\documentclass[final,12pt]{elsarticle}

\usepackage{amssymb}

\journal{Chemical Physics Letters}

\begin{document}

\begin{frontmatter}



\title{Enhanced quantum entanglement in the non-Markovian dynamics of 
biomolecular excitons}


\author[label1,label2]{M. Thorwart}
\author[label2]{J. Eckel}
\author[label3]{J.H. Reina}
\author[label1]{P. Nalbach}
\author[label2,label4]{S.\ Weiss}

\address[label1]{Freiburg Institute for Advanced Studies (FRIAS), Universit\"at Freiburg,
79104 Freiburg, Germany}
\address[label2]{Institut f\"ur Theoretische Physik, Heinrich-Heine-Universit\"at D\"usseldorf,
40225  D\"usseldorf, Germany}
\address[label3]{Departamento de F\'isica, Universidad del Valle, A.A. 25360, Cali, Colombia}
\address[label4]{Niels-Bohr-Institute, Universitetsparken 5, 2100 Copenhagen, Denmark}

\begin{abstract}
We show that quantum coherence of biomolecular excitons is maintained over exceedingly long times due to the  constructive role of 
their non-Markovian protein-solvent  environment.  Using a numerically exact approach, we demonstrate 
 that a slow quantum bath helps to sustain quantum entanglement of two pairs of F\"orster coupled excitons, in contrast to a  
Markovian environment. We consider the crossover from a 
fast to a  slow bath and from weak to strong dissipation and  show that a slow bath can generate robust entanglement.
This persists to surprisingly high temperatures, 
even higher than the excitonic gap and  is absent for a Markovian bath. 
\end{abstract}


\end{frontmatter}


\section{Introduction}
\label{introduction}
Quantum coherent dynamics at the initial stages of 
photosynthesis in complex biomolecular structures seems to promote the efficiency of 
energy transfer from the light-harvesting antenna complexes to the chemical reaction 
centers  \cite{Engel07,Lee07,Brixner05,Herek02}. This hypothesis has  recently been boosted by experiments revealing long-lived quantum coherent excitonic dynamics in the energy transfer  among bacteriochlorophyll complexes over a surprisingly long time of around 600 fs measured at $77$ K \cite{Engel07}. 
 In addition, electronic coherences between 
excited states  in  purple photosynthetic bacteria have been monitored in a two-color photon echo experiment \cite{Lee07}.
 Both
works lead to the conclusion that the collective long-range
electrostatic response of the biomolecular protein environment to the electronic
excitations should be responsible for the long-lived quantum
coherence. 
Furthermore, the obtained time scales \cite{Lee07} for the short-time 
dynamics of the nuclear modes
coupled to the excitonic states of two chromophores 
are almost identical. This points to the special and 
constructive role of the quantum environment for the photoexcitations.  The often assumed coupling of
the chromophores to fast  and independent quantum baths does
not hold in this case. In fact, the two chromophores are embedded in the {\em same\/} protein-solvent environment. 
These results corroborate experimental studies \cite{Brixner05} which show that energy transport sensitively depends on the  spatial properties of the delocalized excited-state wave functions of the {\em whole\/} pigment-protein complex. 
In addition,  there are reports of coherently controlled wave packet quantum dynamics artificially generated by laser
pulses in the light-harvesting antenna 
 of the bacteria {\em Rhodopseudomonas acidophila\/}
\cite{Herek02}.   

An appropriate theoretical description of the biomolecular quantum dynamics 
has to account for  the environmental time scales  typically being of the same order of magnitude as the excitonic time scales \cite{Gilmore06}. This fact renders the 
dynamics highly non-Markovian and 
rather elaborate techniques are required to address the entire cross-over from fast to slowly responding baths.  

Here, we perform a deterministic evaluation of real-time path integrals  \cite{QUAPI1,QUAPI2,Thorwart98}  where all non-Markovian effects are exactly included. We provide numerically exact results for the quantum coherent 
dynamics of photoexcitations in coupled chromophores, where the
time evolution of the protein-solvent bath happens on time scales comparable to the exciton dynamics.  
We show that the non-Markovian effects help to sustain quantum coherence over rather long times. 
Furthermore, quantum entanglement 
 of two chromophore pairs is shown to be more
 stable under the influence of a non-Markovian bath.  
 Even at high temperatures, 
a slow bath can generate  considerable 
 entanglement, a feature   absent in the Markovian case.
We mention that recently, quantum 
entanglement of two optical two-level systems coupled to a
common localized environmental mode has been studied beyond the Markov approximation 
at zero temperature \cite{Bellomo07}.

\section{Model for a single chromophore pair}
\label{singlepair}
A single chromophore (index $i$) is modeled as a quantum
two level system described  by Pauli matrices 
$\tau^i_{x,y,z}$ with energy gap $E_i$ between ground-state 
$|g_i\rangle$ and excited state $|e_i\rangle$ \cite{Gilmore06}. 
The protein-solvent environment 
is formalized as a harmonic bath yielding the standard 
spin-boson Hamiltonian for each chromophore \cite{Gilmore06}. 
The F\"orster coupling between two chromophores is given by  
 $H_{12}=\frac{\hbar\Delta}{2} (\tau_x^1 \tau_x^2 + 
\tau_y^1 \tau_y^2)$~\cite{Gilmore06}. Observing that the 
fluorescence lifetime of the chromophores is much larger 
than all other time scales, we may neglect the radiative decay \cite{Gilmore06}. 
Then, the system's total 
number of excitations is a constant of motion, and the two-chromophore system can be
 reduced to a single spin-boson model of one chromophore 
pair (with the Pauli matrices $\sigma_{x,z}$):
\begin{equation}
H_{1}= \frac{\hbar\epsilon }{2}\sigma_z + \frac{\hbar\Delta}{2} \sigma_x 
+ \hbar\sigma_z  \sum_\kappa \lambda_\kappa (b^\dagger_\kappa + b_\kappa) + 
\sum_\kappa \hbar \omega_\kappa b^\dagger_\kappa  b_\kappa \, ,
\end{equation}
where $\epsilon =E_1-E_2$, and  $b_\kappa$  describe 
bosonic bath operators with couplings $\lambda_\kappa$. We consider equal
chromophores $E_1=E_2$; the effective basis for a chromophore pair 
is given by $\{{|\uparrow \rangle} = {|e_1 g_2 \rangle}, 
{|\downarrow \rangle} = {|g_1 e_2 \rangle} \}$. 

The spectral density \cite{Weiss} follows from a microscopic derivation
\cite{Gilmore08}. Different forms of a 
Debye dielectric can be assumed, but in any case, lead to an 
Ohmic form,  
$G(\omega) = 2 \pi  \alpha \omega e^{-\omega/\omega_c}$. 
The dimensionless  damping  constant $\alpha$ of the protein-solvent 
can be related to the parameters of the dielectric model \cite{Gilmore08}. 
One finds for the order of magnitude of
 $\alpha \sim 0.01-0.1$ \cite{Gilmore06,Gilmore08}.  
We use an exponential form of the 
cut-off at frequency $\omega_c$. This sets the
time-scale for the bath dynamics  and is related to
the reorganization energy $E_r \sim 2\alpha \hbar \omega_c$. If
$\Delta \ll \omega_c$ and $\alpha \ll 1$, 
the bath evolves fast compared to the
system and loses its memory quickly, rendering a Markovian
approximation and the standard Bloch-Redfield
description \cite{may} suitable. This situation is ubiquitous 
in many physical systems, e.g., ion traps, quantum dots, 
and superconducting devices \cite{Shnirman}. It is qualitatively 
different for the excitons in the biomolecular environment. Here,   
$\hbar \omega_c$ is  typically of the order of  $\sim 2-8$ 
meV, while the F\"orster
coupling constants can range from $\hbar\Delta \sim  0.2$ meV$-100$
meV \cite{Gilmore06,Gilmore08}. 
Hence, the bath responds slower than the dynamics of the
excitons evolves, and non-Markovian effects become 
dominant.
Coherent oscillations in a strongly damped two-state system with
$\alpha > 1$ and $\Delta  \gtrsim \omega_c$ have been found using  
 quantum Monte Carlo simulations \cite{Egger94,Muhlbacher03} 
and by applying the numerical renormalization group \cite{Bulla05}. 

\begin{figure}[t]
\begin{center}
\includegraphics[width=0.5\textwidth]{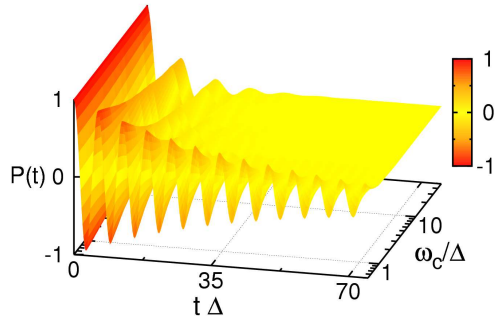}
\caption{ \label{fig1}  Population difference $P(t)$ for a single
chromophore pair and full cross-over from a Markovian 
to a non-Markovian bath.  Parameters are 
$k_BT=0.1\hbar\Delta$, and $\alpha=0.1$. } 
\end{center}
\end{figure}

\section{Population dynamics of a single chromophore pair}
\label{population}
Here, we use the quasiadiabatic
 path-integral (QUAPI) \cite{QUAPI1,QUAPI2,Thorwart98} to calculate 
the population difference  
$P(t) = \langle \sigma_z \rangle_t$ \cite{Weiss} in the  single pair. 
We choose $P(0)=1$. Figure \ref{fig1} shows the results for
$\alpha=0.1$ (parameters correspond to {\em measured\/} 
values summarized in Ref.\ \cite{Gilmore06}). 
$P(t)$ decays with time in an oscillatory way. The decay
occurs faster for large $\omega_c$ while for small $\omega_c$, the
slow bath sustains  more coherent  oscillations. In general, 
 for smaller $\omega_c$ the spectral weight of the
bath modes around the system frequency $\Delta$ is  suppressed and
the decohering influence is reduced, yielding prolonged coherence. 
In fact, choosing $\alpha=0.1, \omega_c=0.1 \Delta$ (consistent with 
\cite{Engel07}), 
we find a coherence  time of $~1$ ps which agrees well with the measured value 
of at least $660$ fs \cite{Engel07}, given the complexity of the setup. 
In passing, we note that we have compared with the Born-Markov result \cite{Weiss} for 
$P(t)$ for small $\omega_c$ and found strong disagreement, as expected (not shown). 

\section{Model for two coupled chromophore pairs}
\label{chromophorepair}
Next, we address entanglement between two chromophore pairs under the
influence of a slow bath. We consider two equal pairs
described by  $\sigma_{x/z,i}$, 
coupled by an interpair F\"orster interaction $J$ and coupled to a harmonic bath.  The total Hamiltonian reads
\begin{eqnarray}\label{ham2}
H_2 & = & \sum_{i=1,2}\frac{\hbar\Delta}{2} \sigma_{x,i} + 
\hbar J(\sigma_{x,1}\sigma_{x,2} + 
\sigma_{y,1}\sigma_{y,2})   \\
& &  +\frac{\hbar}{2} (\sigma_{z,1}+\sigma_{z,2})
\sum_\kappa {c}_\kappa (b^\dagger_\kappa + b_\kappa) + 
\sum_\kappa \hbar \omega_\kappa b^\dagger_\kappa  b_\kappa   ,
\nonumber 
\end{eqnarray}
whose basis refers to the states $\{ {|\uparrow_1 \rangle} = {|e_1 g_2 \rangle}, 
{|\downarrow_1 \rangle} = {|g_1 e_2 \rangle}, {|\uparrow_2 \rangle} = {|e_3 g_4 \rangle}, 
{|\downarrow_2 \rangle} = {|g_3 e_4 \rangle }\}$. 
As before, the bath spectral density 
follows from a Debye dielectric model, again yielding   the Ohmic form.
The time-dependent reduced density matrix $\rho(t)$ is computed 
using an adapted QUAPI scheme. 
%
\begin{figure}[t]
\begin{center}
\includegraphics[width=0.5\textwidth]{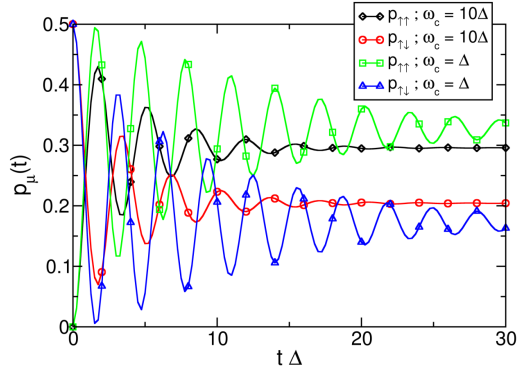}
\end{center}
\caption{ \label{fig2} Time evolution of the populations $p_\mu(t)$ for two coupled chromophore pairs for a slow ($\omega_c= \Delta$) and a fast ($\omega_c= 10 \Delta$) bath, for  $k_BT=0.1\hbar\Delta$, $\alpha=0.1$, and $J=0.1\Delta$.} 
\end{figure}
Figure \ref{fig2} shows the
time-evolution of the populations $p_{\uparrow\uparrow}(t)=
p_{\downarrow\downarrow}(t)$ and 
$p_{\uparrow\downarrow}(t)=p_{\downarrow\uparrow}(t)$ 
of the four basis states for different values of $\omega_c$ for the initial
preparation ${|\psi_0\rangle}=({|\uparrow_1\downarrow_2\rangle} + 
{|\downarrow_1\uparrow_2\rangle})/\sqrt{2}$. After a transient
oscillatory behavior, the stationary equilibrium values are reached.
The corresponding decay occurs on shorter times for large 
$\omega_c$, i.e., fast baths, compared to the rather slow decay for
small  $\omega_c$. 

\section{Entanglement of two chromophore pairs}
\label{pairentanglement}
To quantify the two-pair quantum correlations, we study the 
entanglement along the negativity $N(t)=\max\{0, -2 \zeta_{\rm min}(t) \}$ \cite{Peres,Horo}, 
where $\zeta_{\rm min}(t)$ denotes the smallest
eigenvalue of the partially transposed reduced density operator with the matrix elements $\rho_{m\mu,n\nu}^{T_2}= \rho_{n\mu,m\nu}$.  A separable state has $N=0$, while for a
maximally entangled state, $N=1$.

\begin{figure}[t]
\begin{center}
\includegraphics[width=0.5\textwidth]{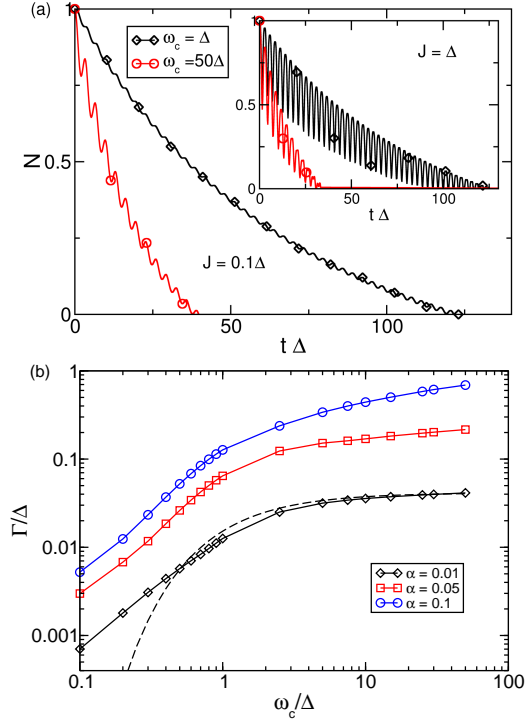} \\
\includegraphics[width=0.5\textwidth]{fig3b}
\end{center}
\caption{ \label{fig3} (a) Time evolution of the negativity $N(t)$
for the cut-off $\omega_c= \Delta$ (black) and $\omega_c= 50
\Delta$ (red) for $\alpha=0.01$ and for  $J=0.1 \Delta$
(main) and $J=\Delta$ (inset). Moreover, $k_BT=0.1 \hbar \Delta$. 
(b) Decay constant $\Gamma$ as a
function of the cut-off $\omega_c$ for different values of
$\alpha$ (symbols with solid lines) for $k_B T=0.1\hbar \Delta, J=0.1\Delta$. Dashed line: 
Corresponding one-phonon result $\Gamma =\Gamma_0 e^{-\Delta/\omega_c}$, where the proportionality 
constant $\Gamma_0=0.041 \Delta$ has been obtained from a fit to the three data points  $\omega_c=25 \Delta, 
30 \Delta, 50 \Delta$. 
} 
\end{figure}

Figure \ref{fig3}(a) shows $N(t)$ for two 
values $\omega_c=\Delta$, and $\omega_c= 50 \Delta$, for the maximally
entangled initial state $|\psi_0\rangle$. Starting from $N(0)=1$ we
observe a decay to zero with small oscillations superimposed. For the
Markovian bath $\omega_c=50 \Delta$, the decay occurs faster than
for the non-Markovian bath $\omega_c= \Delta$, indicating 
that entanglement survives on a longer time scale for the slow
bath. For a larger interpair
coupling $J=\Delta$, the superimposed oscillations are more
pronounced (Fig.~ \ref{fig3}(a) inset) which is due to constructive
interference of the transitions within the  chromophore.  
For a quantitative picture, we
fit an exponential $N(t)=N_0 \exp(-\Gamma t) + N_1$ with a 
constant $\Gamma$ which contains the influence of the bath. 
Figure \ref{fig3}(b) shows the dependence of $\Gamma$ on $\omega_c$.
Clearly, $\Gamma$ strongly decreases 
for small $\omega_c$, while for large $\omega_c$, the decay rate saturates to a constant value. 
The dependence of $\Gamma$ on $\omega_c$ is more pronounced for
larger values of $\alpha$. This observation indicates that 
entanglement could be at least as robust 
in biomolecular systems as in other macroscopic 
 condensed-matter devices \cite{Shnirman} which display quantum coherent behavior.
We note that this tendency is already captured by the result of a one-phonon perturbative analysis, i.e., 
in second order perturbation theory. It would predict that for $\omega_c \le \Delta$ no bath modes for 
efficient one-phonon processes are available and $\Gamma \propto G(\Delta) \propto e^{-\Delta/\omega_c}$. A fit 
to the data points in the region of large $\omega_c$ is shown by the dashed 
line in Fig.\  \ref{fig3}(b) for $\alpha=0.01$. Clearly, when $\omega_c \le \Delta$, multi-phonon 
contributions become significant. This is even more pronounced for larger $\alpha$ (not shown). 

\begin{figure}[t]
\begin{center}
 \includegraphics[width=0.5\textwidth]{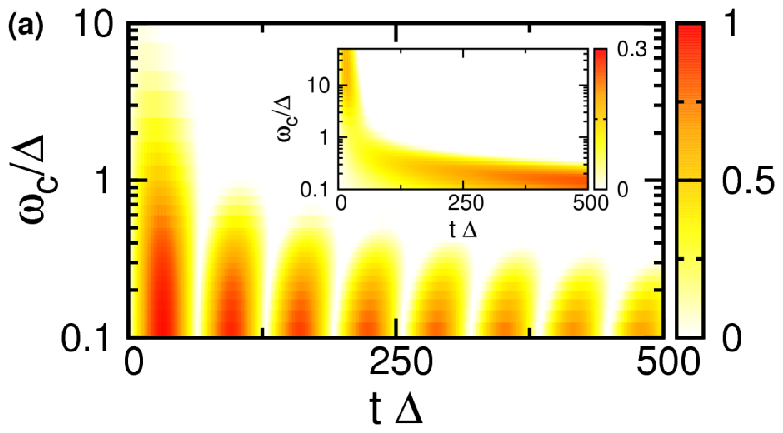} \\
\includegraphics[width=0.5\textwidth]{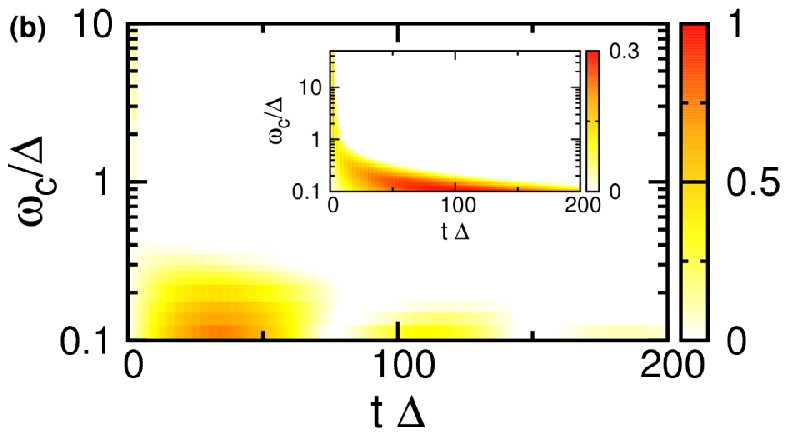}
\end{center}
\caption{ \label{fig4} (a) Negativity $N(t)$
as a function of $\omega_c$ for 
$J=0.1 \Delta$
(main) and $J=0$ (inset), for $\alpha=0.01$ and $k_BT=0.1 \hbar\Delta$. (b) Same as in (a), but  for the strong coupling case  $\alpha=0.1$. } 
 \end{figure}

 To study the cross-over between fast and slow baths,
 we show $N(t)$ for varying $\omega_c$ in Fig.~\ref{fig4} 
 for the 
 initial state 
 ${|\psi_1\rangle} = {|\uparrow_1 \uparrow_2 \rangle}$.   Figure \ref{fig4}(a) shows the result for $J=0.1
 \Delta$. 
 The entanglement is rather quickly destroyed in the regime
 $\omega_c\gg \Delta$. On the other hand, we find a regular
 oscillatory decay for $0.1\Delta  \lesssim \omega_c \lesssim \Delta$.
 In this regime, complete entanglement  disappearance and  
 revivals alternate. The time scale of the entanglement oscillations
 is given by $2\pi/J$. The constructive role of a slow bath is
 further illustrated in  the inset of Fig.~\ref{fig4}(a), where 
 $N(t)$ is shown for 
  $J=0$. In fact, in the regime $\omega_c < \Delta$, we find that
 entanglement between the two pairs is generated by their common 
 interaction with a slow bath. Most interestingly, for
 $\omega_c=0.1 \Delta$, $N(t)$  steadily grows even over rather long
 times up to $t \Delta=500$. In view of the single-pair results
 described above, this seems counterintuitive since for small
 $\omega_c$, a reduced influence of the bath modes would be expected.
 However, in this regime, the bath is rather efficient in generating
 entanglement. This feature  survives even for larger values of  
 $\alpha$, see Fig.~\ref{fig4}(b). The oscillatory
 behavior of the entanglement generation still occurs for
 $J=0.1\Delta$, where $N(t)$ assumes all values between zero and one.
 The bath-induced destruction happens here earlier due to the large
 $\alpha$.  Entanglement is also produced when $J=0$, see inset of 
 Fig.~\ref{fig4}(b), for $0.1\Delta \lesssim \omega_c \lesssim
 \Delta$. Also here, $N(t)$  can even reach the maximal value at
 intermediate times.   

The generation of entanglement can be qualitatively understood by performing a 
polaron-like transformation $U=\exp[i(\sigma_{z,1}+\sigma_{z,2})p/2]$ with 
$p=\sum_\kappa ic_\kappa (b_\kappa^\dagger - b_\kappa)$. 
Setting $J=0$ and assuming that the two qubits are spatially close enough,  
the resulting Hamiltonian $\tilde{H}_2=U^\dagger H_2 U$ acquires 
an effective direct coupling $\tilde{H}_{2,\rm int}=\hbar J_{\rm eff} 
\sigma_{z,1}\sigma_{z,2}$ with $J_{\rm eff}=-\alpha \omega_c/8\sim E_r$ 
(note that restrictions to the applicability of $U$ apply. Details will be published 
elsewhere). 
 Then, the  long-wave length bath modes are efficient in generating  
coherent coupling, and thus entanglement. Its dynamical 
generation occurs on a time scale $1/J_{\rm eff}$, see insets
of Fig.~\ref{fig4}. On the other hand, damping destroys  
coherence on a time scale $\Gamma$ related to $\omega_c$, see Fig.~\ref{fig3},
i.e.,  
 for large $\omega_c$, damping beats entanglement generation. 

\begin{figure}[t!]
\begin{center}
 \includegraphics[width=0.5\textwidth,keepaspectratio=true]{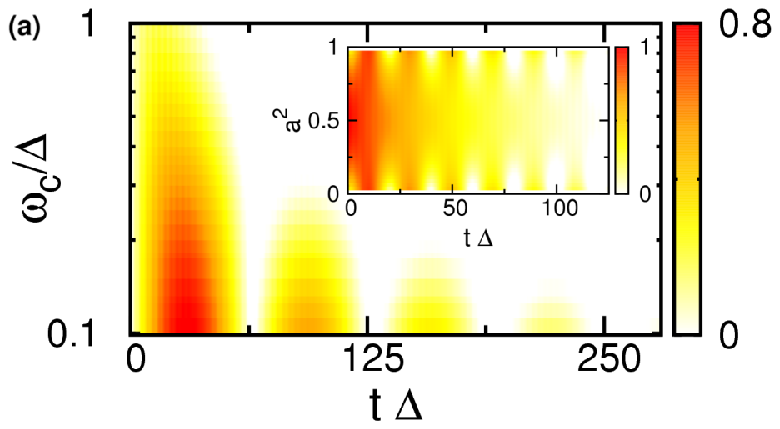} \\
\includegraphics[width=0.5\textwidth,keepaspectratio=true]{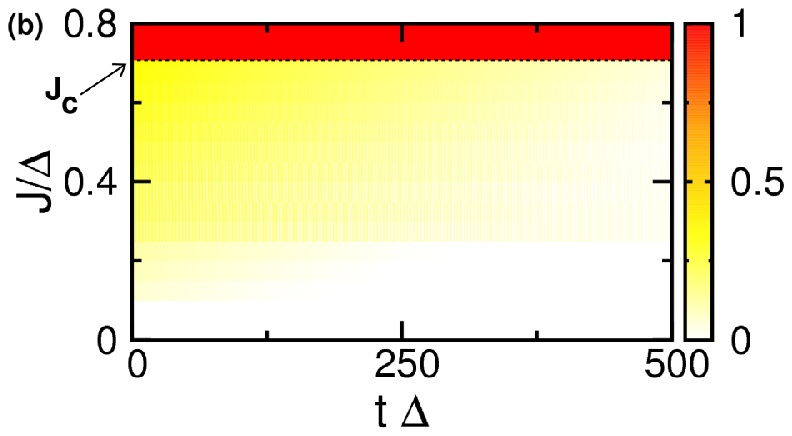} 
\end{center}
\caption{ \label{fig5}
(a)   Negativity $N(t)$ for varying $\omega_c$;  $J=0.1\Delta,
\alpha=0.01$ and $k_BT=\hbar\Delta$ (Main). Inset: $N(t)$ for different initial  preparations ${|\psi_2\rangle}=a{|\uparrow_1\downarrow_2\rangle} + 
b {|\downarrow_1\uparrow_2\rangle}$, for $J=0.1\Delta, \alpha=0.01,
k_BT=0.1\hbar\Delta$, and $\omega_c=\Delta$.  (b) $N(t)$ for varying 
$J$; $\alpha=0.01, k_BT=0.1\hbar\Delta, \omega_c=0.1\Delta$. $J_c=1/\sqrt{2}$ marks the border  above which the initial ground state belongs to a DFS. }
\end{figure}
So far, we have studied 
not so high temperatures, similar to the
experimental conditions in Refs.~\cite{Engel07,Lee07}. However, 
in Fig. \ref{fig5}(a) (main) we plot $N(t)$ for varying $\omega_c$, for $k_BT=\hbar\Delta$, for the initial state $|\psi_1\rangle$.
We still find large entanglement oscillations at short to intermediate times, for $0.1\Delta \lesssim \omega_c\lesssim \Delta$ despite the rather large temperature: this is an outstanding hardware feature that could 
provide a useful resource for the artificial design of controlled, robust, 
 and efficient biomolecular nanostructures for quantum information processing 
\cite{Kroutvar04,Rabi06,Thorwart02}.  

Furthermore, we have varied the initial preparation to the state  
${|\psi_2\rangle}=a{|\uparrow_1\downarrow_2\rangle} + 
b {|\downarrow_1\uparrow_2\rangle}$ with $a^2+b^2=1$. The inset of   
Fig.~\ref{fig5}(a) shows $N(t)$ for varying $a^2$ and $J=0.1\Delta$. 
${|\psi_2\rangle}$ is maximally entangled for $a^2=1/2$, for which $N(t)$ decays monotonously with time, while away from this
region the negativity  again shows collapses and revivals. For the borders $a^2\to 0,1$, ${|\psi_2\rangle}$ is a separable state, but
entanglement is rather quickly generated with time before it finally 
dies out. Robust entanglement thus depends on the  initial preparation and is favored by the choice of initially  separable (or weakly entangled) states.

Finally, we analyze the dependence on the interpair coupling $J$. 
The negativity $N(t)$ is shown in Fig.~\ref{fig5}(b) for varying $J$ for the respective ground state as the initial preparation.   From Eq.\   (\ref{ham2}) it follows that a critical value $J_c=1/\sqrt{2}$ exists such that for $J\ge J_c$, the state 
$|\psi_g\rangle=({|\uparrow_1\downarrow_2\rangle} -  
 {|\downarrow_1\uparrow_2\rangle})/\sqrt{2}$ is the two-pair groundstate, which, however, belongs 
to a decoherence-free subspace (DFS) of $H_2$
 \cite{Reina02}. Hence, $N(t)$ remains constantly maximal. For $J < J_c$, the ground state has some weight outside of the DFS and 
hence suffers from decoherence.  

We emphasize that we have formulated our approach in quantum statistical terms. It directly involves the reduced density operator which describes mixed ensemble states. Thus, the entanglement dynamics reported here manifests itself in a statistical 
many-particle ensemble and, hence, would appropriately allow the design and implementation of 
 robust biomolecular entanglement proof-of-principle experiments with current technology.

\section{Conclusions}

In this Letter, we have shown that non-Markovian effects are vital for  the correct description of 
the entanglement dynamics of biomolecular excitonic qubits. 
Recent results obtained from a reduced hierarchy equation approach 
\cite{Fleming09}  are within the lines of our findings. 
 While our QUAPI technique yields 
 numerically exact results, the approach developed in \cite{Fleming09} allows to include non-Markovian corrections 
only up to a certain extent. In either case, the results yield that the validity of the Born-Markov approximation 
for the treatment of slow protein-solvent environments in photosynthetic complexes 
\cite{Aspuru08} is questionable, even in the regime of high temperature; see, e.g., our Fig.\ 
\ref{fig5}. 

Our results are relevant to molecular    
architectures~\cite{Crooker02,Reina04} and ultrafast processes 
\cite{atto}.  
They could prove crucial in the design of artificial 
 light harvesters
for robust biomolecular entanglement, with 
enhanced energy transfer rates~\cite{Becker06} for the  
control~\cite{Thorwart02} of quantum bits. Most importantly, our predictions could be tested with currently 
available experimental techniques 
\cite{Engel07,Lee07,Brixner05,Herek02}. For instance, advanced ultrafast spectroscopy techniques 
allow to measure the time dependence of the elements of the two-particle reduced density matrix from which an entanglement measure 
can be derived straightforwardly. Similarly, such an experiment would give direct insight into the interpair coupling strength which can directly be derived from the oscillation period of the entanglement measure. It should also be possible to prepare desired initial 
states. A further crucial topic is the behavior of this effect at higher temperature which can be experimentally doable. We hope that our results can be experimentally demonstrated in the near future.

\appendix
\section{Acknowledgment}
We acknowledge support by the DFG SPP 1243, COLCIENCIAS grs. 1106-14-17903, 1106-452-21296, 
DAAD PROCOL, and the Excellence Initiative of the German Federal and State Governments and  CPU time from 
the ZIM in D\"usseldorf.




\begin{thebibliography}{00}
\bibitem{Engel07} G.S. Engel, 
T.R. Calhoun, E.L. Read, T.-K. Ahn, T. Mancal, Y.-C. Cheng, R.E. Blankenship, G.R. Fleming,  
Nature {\bf 446} (2007) 782. 

\bibitem{Lee07} H. Lee, Y.-C. Cheng, G. R. Fleming, 
Science {\bf 316} (2007) 1462. 

\bibitem{Brixner05} T.  Brixner, 
 J. Stenger, H.M. Vaswani, M. Cho, R.E.
Blankenship, G.R. Fleming, 
Nature {\bf 434} (2005) 625. 

\bibitem{Herek02}J. L. Herek, 
 W. Wohlleben, R. J. Cogdell, D. Zeidler, M. Motzkus, 
Nature {\bf 417} (2002) 533. 

\bibitem{Gilmore06}  J. Gilmore, R. McKenzie, 
Chem. Phys. Lett. {\bf 421} (2006)  266.

\bibitem{QUAPI1}
N. Makri, 
J. Math. Phys. {\bf 36} (1995) 2430.

\bibitem{QUAPI2}
N. Makri, 
 E. Sim, D. Makarov, M. Topaler,
Proc. Natl. Acad. Sci. USA {\bf 93} (1996) 3926. 

\bibitem{Thorwart98}
M. Thorwart, 
P. Reimann, P. Jung, R. F. Fox, 
Chem. Phys. {\bf 235} (1998) 61.

\bibitem{Bellomo07} B. Bellomo, R. Lo Franco, G. Compagno,  
Phys. Rev. Lett.  {\bf 99} (2007) 160502.

\bibitem{Weiss} U. Weiss,  Quantum Dissipative Systems, 3rd ed., World Scientific, Singapore, 2008.
 
 \bibitem{Gilmore08} J. Gilmore, R. McKenzie, 
 J. Phys. Chem. A {\bf 112} (2008) 2162.

\bibitem{may}
V. May, O. K\"uhn, Charge and energy transfer
dynamics in molecular systems, Wiley, Berlin, 2001. 

\bibitem{Shnirman}Y. Makhlin, G. Sch\"on, A. Shnirman, 
 Rev. Mod. Phys. {\bf 73} (2001) 357. 

\bibitem{Egger94} R. Egger, C. H. Mak, 
 Phys.\ Rev.\ B {\bf 50} (1994) 15210.

\bibitem{Muhlbacher03}L. M\"uhlbacher, R. Egger, 
 J.\ Chem.\ Phys.  {\bf 118} (2003) 179.

\bibitem{Bulla05} R. Bulla, 
H.-J. Lee,  N.-H. Tong, M. Vojta, 
Phys.\ Rev.\ B {\bf 71} (2005) 045122.

\bibitem{Peres}A. Peres, 
Phys. Rev. Lett.  {\bf 77} (1996) 1413. 

\bibitem{Horo} M. Horodecki, P. Horodecki, R. Horodecki, 
 Phys. Lett. A {\bf 223} (1996) 1. 


 \bibitem{Kroutvar04}
M. Kroutvar, 
Y. Ducommun, D. Heiss, M. Bichler, D. Schuh, G. Abstreiter,
 J. Finley, 
Nature  {\bf 432} (2004) 81.
%

\bibitem{Rabi06} P. Rabl, 
D. DeMille, J. Doyle, M. Lukin, R. Schoelkopf, P. Zoller, 
Phys. Rev. Lett.  {\bf 97} (2006) 033003.

\bibitem{Thorwart02}
M. Thorwart, P. H\"anggi, 
Phys. Rev. A {\bf 65} (2002) 012309. 

\bibitem{Reina02}J. H. Reina, L. Quiroga, N. F. Johnson,  
Phys. Rev. A {\bf 65} (2002) 032326.

\bibitem{Fleming09}A.\ Ishizaki, G. R. Fleming, J.\ Chem.\ Phys.\ 
{\bf 130} (2009) 234111.

\bibitem{Aspuru08}M.\ Mohseni, P.\ Rebentrost, S.\ Lloyd, and A.\ Aspuru-Guzik, 
J.\ Chem.\ Phys.\ {\bf 129} (2008) 174106.

\bibitem{Crooker02} S. A. Crooker, 
J. A. Hollingsworth, S. Tretiak, V. I. Klimov,  
Phys. Rev. Lett. {\bf 89} (2002) 186802.

\bibitem{Reina04} J. H. Reina,
R.G. Beausoleil,  T.P. Spiller, W.J. Munro, 
Phys. Rev. Lett. {\bf 93} (2004) 250501.

\bibitem{atto}P. Agostini, L.F. DiMauro, Rep.\ Prog.\ Phys.\ {\bf 67} (2004) 813. 

\bibitem{Becker06}  K. Becker, 
J.M. Lupton, J. M\"uller, A.L. Rogach, D.V. Talapin, H. Weller, J. Feldmann, 
Nature Mater. {\bf 5} (2006) 777.

\end{thebibliography}
\end{document}